\documentstyle[12pt]{article}
\begin{document}
\bibliographystyle{unsrt}
\title{ World Wide Student Laboratory}
\author{ Description of Project\\
{}\\
Anatoli Arodzero\\
Department of Physics, University of Oregon\\
Eugene, Oregon 97403, USA}
\date{February 20, 1995\\
Revised June 19, 1998 \thanks{Copyright \copyright 1995-1998 A. Arodzero}}
\maketitle
\section{Introduction}

\hspace{5mm} This paper describes a new framework in educational technology - the 
World Wide Student Laboratory Project. Its main goals are to  improve the efficiency of educating students in the principles of advanced experimental research, to stimulate student's interest in science and to 
provide professional resources  to educators.\\
\indent
        The World Wide Student Laboratory ( WWSL ) Project expands upon traditional laboratory study. 
It allows students to study natural phenomena while  learning
 an Internet-based  telecommunication and computer technology and studying the methods of 
transmission, processing and  analyzing  experimental data.\\
\indent
In this publication, the concepts of the World Wide Student Laboratory are outlined, a possible idea for structure is discussed and examples of various laboratory studies (projects) are shown. 

\section{Concept of the World Wide Student Laboratory}

\hspace{5mm}Based on the Internet, the World Wide Web (WWW) has been
rapidly growing over the last several years.  The WWW
is a combination of a large number of computer servers which can
send documents or "pages" to Internet clients using a
Web-browser software for navigation between the different "sites."\\
\indent
From the beginning the WWW has primarily been used for educational purposes in three ways \cite{1}:\\
\begin{quotation}
\indent
 - to provide students with a wider access to information;\\
\indent
 - as a communications tool supplementing traditional forms of education, raising
the efficiency of interaction between the educators and students;\\
\indent
 - as "a virtual classroom", " a virtual laboratory", as an
integrated interface for distance learning.\\
\end{quotation}
\indent
The student's laboratory  practice work is the main form
of experimental research study. The efficiency of traditional labs have certain 
limits:\\
\begin{quotation}
\indent
  -often students do not have  enough opportunities to deeply explore a
 number of important sections of laboratory courses . Not all universities can support a wide spectrum of
laboratory research setups because of their high price, service
cost, the lack of necessary space, etc.;\\
\indent
  -more often laboratory classes  are too short to obtain persuasive
research results of the given process or phenomenon;\\
\indent
  -in many cases, the laboratory instruction and student research projects  do not 
fully reflect the current level of professional experimental research.\\
\end{quotation}
\indent
In the framework of the WWSL project, we offer  new methods of
educational technology which complement the traditional methods and upgrade the 
standard of experimental research study even with the use of existing lab equipment. 
These methods may  also significantly reduce the above mentioned limits of laboratory
practice work.\\
\indent
        The concept of  WWSL  may be expressed in the following way:\\
\begin{quotation}
{\it
\indent
	World Wide Student Laboratory (WWSL) is a dynamic international Internet-based collaboration of resources from the educational and research labs of universities, institutions, government research centers and companies.  The main element of WWSL is the educational or research setup. This setup has an interactive connection through the Internet with a particular thematic WWSL-Site (Research Group Site). Through the Research Group Sites, students, under the instruction of an educator, are able to control experiments, take, process and analyse data, regardless of their location.  The results obtained by each student are available to all other students who are participants of the particular research group and can be used in their own projects.}
\footnote{The idea of such a organisation of lab studies came to me in 1987, when I 
was  project manager of development of the new Department of Fundamental Science at the 
Moscow State Technical University. But the Internet at that time was weakly developed.}
\end{quotation}

The main features of the WWSL project involve the following:\\
\begin{quotation}
\indent
-the possibility of studying and demonstrating certain effects
and phenomena unobservable in traditional lab studies. For example, the
study of the effects which require simultaneous tests in
different geographical points of the world, in different
environments, during long time intervals, etc.;\\
\indent
-as a result of unification of the experimental data obtained
from a network of setups, it is possible to
study "subtle" processes and/or processes which require a large volume
of data. This allows  exploration of the origins of
instrumental or methodological errors and  demonstrates methods of
extracting information from noisy data;\\
\indent
-it is possible to run certain experimental projects in parallel with 
mathematical modelling. Integrating the two allows more global understanding of the 
phenomena;\\
\indent
-any student who has  access to the Internet, regardless of
their location, may become a member of the WWSL. This also makes the WWSL a perfect 
tool for distance education;\\
\indent
-students have around the clock access to experimental set-ups, and the ability to work on their own schedule;\\
\indent
-educators have the possibility to use the data of experiments on-line 
through the WWSL for lectures;\\
\indent
-by using the WWSL, students can have access to the real data from
"professional" scientific experiments, and educators may use this "live" data for 
educational purposes;\\
\indent
-the use by WWSL of  well known elements of the Internet such as chat rooms, white-boards, etc. can allow students and educators to have
"global seminars " for each experiment;\\
\indent
-individual student projects can be integrated in the form of a final
report of the research work. The participation in making such a report will teach 
the student the ethics of joint research, increase motivation and give significance to their work;\\
\indent
-setting up new WWSL software modules may well be an attractive and challenging 
project for students in computer science;\\
\indent
-members of WWSL may not only be students from universities but also from colleges and high schools.\\
\end{quotation}
\hspace{5mm}
\indent
WWSL should enrich the personal
research results of each student and help lab studies approach the top-level scientific research. {\bf It is important to
emphasise that WWSL is not a "virtual" laboratory. It is a real lab where actual  
experiments (with all real-world ``noisy'' effects) can be made.}\\

\indent
The traditional form of conducting some experiments is still preferable for laboratory studies.  However, from 
our point of view, the following projects would be better served by WWSL:\\
\begin{quotation}
\indent
-projects that require a long time to complete;\\
\indent
-projects which require a certain preliminary theoretical and/or practical 
preparation;\\
\indent
-projects which require repeated change of experimental conditions;\\
\indent
-projects which require special conditions for their fulfilment.\\
\end{quotation}

\section{The Structure of WWSL}

\hspace{5mm}
\indent
Presently, the World Wide Student Laboratory is developing using the following structure:\\  
\indent
The general coordination of WWSL operations is conducted by the {\it WWSL Coordination Center} which fulfils the following functions:\\
\begin{quotation}
\indent
-supports the official Web-site of the WWSL Project;\\
\indent
-registers organizations and individuals which will be participants in the WWSL;\\
\indent
-monitors status of WWSL activity;\\
\indent
-manages the development of methodology of WWSL studies and its data;\\
\indent
-maintains the WWSL-FAQ, chat rooms, etc.\\
\end{quotation}
\indent
A {\it Research Group} Site  would be created to conduct a particular series of student lab studies.  It combines the sites of {\it Experimental Setups} of participating organizations.  If it is required, the activity of each {\it Research Group} can be defined by a {\it Research Program} which is prepared  by {\it Educators} from participating Universities.\\
\indent 
{\it The Research Group} Site has the following functions:\\
\begin{quotation}
\indent
-registers the participants in given lab studies;\\
\indent
-integrates {\it Student's} Sites who are participating in a given project;\\ 
\indent
-provides access to experimental data from all Experimental Setups;\\
\indent
-maintains {\it Research Program} if required, and/or a schedule for taking data;\\
\indent
-maintains the pages with information necessary to complete
these studies;\\
\indent
-maintains the FAQ and chat room page;\\
\indent
-maintains a page with test questions;\\
\indent
-maintains a page for a combined student report.\\
\end{quotation}
\indent
The Site of the {\it Experimental Setup} may be located at a
computer which is a part of the setup and is used for data
storage, data processing and analysis of the results from
that setup.  It is maintained by the {\it Educator} in charge of the
setup.\\
\indent
{\it Students} use sites they must create (analogous to an electronic lab book) for access to WWSL resources, to control experimental setups and for processing and analysing experimental data.  Their site would also be used to prepare the final report of a completed experiment.\\
\indent
The structure of WWSL is not complicated and uses existing technology. Although it can take advanage of then, it does not rely on new technical developments at student or
research laboratories or on the  Internet.\\

\indent
Presently, the first stage of the WWSL development has been completed 
\cite{2}.  An Official Site of the WWSL Project has been created on the University of Oregon Physics Department's web server, in Eugene Oregon, USA. On that server, a Site for the "Cosmic Rays" Research Group has been organised.  It combines two student research setups, one of which is located at the University of Oregon, the other is located at the Department of Physics at N. E. Bauman Moscow State Technical University, Moscow, Russia.\\   
\indent
The next stage is going to involve a number of projects for students who are studying 
General Physics, Radiophysics, Statistical Physics and Biophysics. The following projects are under discussion and/or development:\\
\begin{quotation}
\indent
-lifetime of cosmic ray muons;\\
\indent
-Mossbauer effect;\\
\indent
-Coriolis acceleration;\\
\indent
-nuclear magnetic resonance;\\
\indent
-study of the accuracy of global positioning system;\\
\indent
-study of the correlation between the processes of different biological systems.\\
\end{quotation}
 
\section{Examples of possible WWSL projects}
\hspace{5mm}
\indent
Let us  now consider some possible student projects which can be 
completed in laboratory setups for studying the properties of cosmic rays.  These studies 
are part of the curriculum of physics and engineering majors in many universities 
\cite{3}, \cite{4}.\\
\indent 
The setup consists of two detectors, usually scintillating. An absorber of variable 
thickness (a set of lead or tungsten plates) is sandwiched between them. The signals 
from the detectors are read out to a coincidence unit. The unit selects events due 
to cosmic ray particles passing through both detectors.  The output signals from coincidence units are collected 
by a data acquisition system and a computer.\\
\indent
Student projects which can be pursued using these setups can be divided into 
two groups:\\
\begin{quotation}
\indent
-projects for studying certain properties of cosmic rays, for studying of experimental methods and apparatus of cosmic ray physics and high energy physics;\\
\indent
-projects for studying the properties of random  processes (in this case, the flux of 
particles of cosmic rays is considered as a generator of random processes).\\
\end{quotation}
\indent
The first group may contain the following projects:\\
\begin{quotation}
\indent
-study of the composition of cosmic rays near the earth surface;\\
\indent
-study of the angular distribution of the cosmic ray flux;\\
\indent
-study of the fluctuation of energy loss of charged particles;\\
\indent
-study of the effect of the detector's dead time on the character of the statistical 
distribution of the signals.\\
\end{quotation}
\indent
The second group may contain the following projects:\\
\begin{quotation}
\indent
-properties of Poisson data of random events;\\
\indent
-study of the relationship between binomial, Poisson and
normal distributions;\\
\indent
-test of the hypotheses presented by different statistical distributions;\\
\indent
-the "waiting time paradox" for Poisson processes;\\
\indent
-study of correlations in stochastic processes.\\
\end{quotation}
\indent
The experimental data collected with the setups for the detection of cosmic rays allows 
students to fulfil the studies of all the topics listed above simultaneously, and WWSL provides the tools to study them fully.\\

\end{document}